\documentclass[aps,prl,amsmath,amssymb,superscriptaddress,showkeys,showpacs,twocolumn,floatfix]{revtex4-2}

\usepackage{graphicx}
\usepackage{subfigure}
\usepackage{hyperref}
\usepackage{ulem}

\usepackage{lipsum}

\usepackage{graphicx}
\usepackage{dcolumn}
\usepackage{bm}
\usepackage{amssymb}
\usepackage{epsfig}
\usepackage{slashed}
\usepackage{color}

\usepackage{xcolor}
\usepackage[utf8]{inputenc}

\newcommand{\be}{\begin{equation}}
\newcommand{\ee}{\end{equation}}
\newcommand\bea{\begin{eqnarray}}
\newcommand\eea{\end{eqnarray}}

\begin{document}



\title{A first-order deconfinement phase transition in the early universe and gravitational waves}

\author{Fei Gao }
\email[]{fei.gao@bit.edu.cn}
\affiliation{School of Physics, Beijing Institute of Technology, 100081 Beijing, China}

\author{Sichun Sun }
\email[]{sichunssun@bit.edu.cn}
\affiliation{School of Physics, Beijing Institute of Technology, 100081 Beijing, China}

\author{Graham White}
\email[]{g.a.white@soton.ac.uk}
\affiliation{School of Physics and Astronomy, University of Southampton, Southampton SO17 1BJ, United Kingdom}
\date{\today}

\begin{abstract}
We clarify the conditions of the cosmic quantum chromodynamics (QCD) first-order phase transition in the early universe by carefully distinguishing the chiral and deconfinement phase transitions. While the chiral one with light quarks at zero chemical potential is  unlikely to be first order
based on the recent lattice QCD calculations, the latter one can be naturally extended with one extra rolling scalar to be first order. The argument is also valid for the dark QCD theory with arbitrary $N_c$ with a wide range of phase transition temperatures, which can be from hundreds of MeV up to beyond TeV.  Notably, we derive the general formula for the deconfinement phase transition potential of SU($N_c$) gauge theory characterized by the Polyakov loop. With the effective potential in hand, the gravitational wave spectrum is then determined via the sound shell model,  which then enables us to give for the first time the quantitative analysis of the gravitational wave signals coming from the QCD deconfinement phase transition and awaits the check from future space interferometers.
\end{abstract}

\keywords{cosmic QCD transition, Polyakov loop, Gravitational wave spectrum}

\maketitle

\paragraph{Introduction}

 The early universe can be a great laboratory for many Standard Model(SM) and beyond Standard Model (BSM) phenomena with its high temperature and density. Our observations today offer us an understanding of the cosmological history, whereas many BSM or non-standard processes can still come into play without violating the observational constraints\cite{PhysRevD.98.030001,ParticleDataGroup:2022pth}.

Most of the previous studies have been focused on the electroweak sector, while the possible role of the QCD sector in the early universe is relatively overlooked. One important reason is the non-perturbative nature of QCD which awaits to be explored. This essential nonperturbative property of QCD has been found to bring in abundant phenomena. For instance, there are two types of phase transitions (PTs) in QCD~\cite{Roberts:2000aa,Philipsen:2012nu,Lucini:2012gg,Fukushima:2013rx,Fischer:2018sdj,Dupuis:2020fhh}. The chiral PT has been widely studied as it is related to the dynamical generation of the quark mass. There also exists another PT with the Polyakov loop being the order parameter. This PT is relatively separate from the quark sector as it can take place with pure Yang-Mills theory and is of the first order, for SU(3) theory, and the second order for SU(2) theory~\cite{Braun:2007bx,Fister:2013bh,Fischer:2013eca,Reinosa:2014ooa}. With such an intriguing feature, this PT in the gluon sector may be implemented in the early universe.

In particular, there have not been strong constraints for QCD in cosmological experiments, and hence,  it is still possible that
nature deviates from the SM in the strongly coupled sector at higher temperatures. The cosmological history before the Big Bang nucleosynthesis around 10 MeV also leaves many uncertainties for new physics beyond the Standard Model.
It has been argued that coupling scalar fields to QCD may induce chiral symmetry breaking in the early universe\cite{Ipek:2018lhm,Lu:2022yuc,vonHarling:2017yew,Davoudiasl:2019ugw} 
However, 
 the realization of first order PTs in either the Standard Model QCD or dark QCD~\cite{Bai:2013xga,Schwaller:2015tja,Morgante:2022zvc,Jiang:2015qor} is subtle.
As the recent lattice simulation suggests, the chiral PT of QCD in all massless cases with quark flavor $N_f\leq 6$ is consistent with a second-order transition at zero chemical potential~\cite{Cuteri:2021ikv}. A large quark chemical potential may induce a first-order PT together with a fine-tuning of the lepton asymmetry in the early universe~\cite{Gao:2021nwz,Gao:2023djs}.  In a word, the first-order chiral PT requires strong coupling and also large quark chemical potential.

Here in this letter, we find that contrary to the extreme conditions that are applied to realize the chiral PT,  the first-order PT in the gluon sector for SU($N_c$) with $N_c\geq 3$, characterized by the Polyakov loop, is naturally induced through a scalar field and/or gauge coupling. The key difference from the chiral PT is that this PT only requires an effective mass for the gluon field via the gauge coupling.  Here we focus on the cosmic PT after inflation, before BBN, derive and calculate the general form of PT potential with the background condensate in Polyakov loop for SU($N_c$) with $N_c\geq 3$, and then illustrate several benchmark points for possible gravitational wave detection of this first-order PT. Other related quantities, latent heat, transitional temperature, trace anomaly, etc can also be calculated from the general potential of SU($N_c$) with arbitrary $N_c\geq 3$ we derived here. The potential is very interesting for the study of large $N_c$ gauge theory.

\paragraph*{YM coupling and gluon effective mass}

It has been illustrated that the  PT characterized by the Polyakov loop is strongly related to the effective mass of gluon.  Here we estimate the effective mass of the gluon field through the hard thermal loop approximation as:
\begin{equation}
m_g^2=\frac{4\pi N_c}{9}\alpha(\mu^2)T^2,
\end{equation}

To introduce such a  new Yang-Mills PT, one may directly add a new dark YM sector or one can couple a new scalar field to the YM sector as suggested in Ref.~\cite{Ipek:2018lhm},.
The new scalar field is coupled to the gluon field in the Lagrangian as:

\begin{equation}
{\cal L} \supset -\frac{\phi}{4\Lambda_{gpt}}G_{\mu\nu}G^{\mu\nu}
\end{equation}
The scale $\Lambda_{gpt}$ is associated with new physics dynamics much heavier than the energy scale of the gluon PT.
With this additional term,  the one-loop running of the QCD coupling is modified as:
\begin{equation}
\alpha^{-1}(\mu^2)=\frac{33-2n_f}{12\pi}\mathrm{ln}(\frac{\mu^2}{\Lambda_{QCD}^2})+4\pi\frac{<\varphi>}{\Lambda_{sc}},
\end{equation}
with  $\mu$ the renormalization scale,  $n_f$ the flavor, and $\Lambda_{QCD}$ the intrinsic scale of the theory . In Ref.~\cite{Ipek:2018lhm}, it has been argued that a negative expectation value of the scalar field can be carefully chosen to cancel the SM running so that the full coupling runs into infinity.  Such an infinite coupling will likely induce a  QCD PT. Besides the request for strong coupling,  it is still difficult to induce a first-order chiral PT as aforementioned~\cite{Cuteri:2021ikv}.
{\color{red} 
}

With an effective mass appearing at a certain energy scale, the PT characterized by the Polyakov loop will be naturally induced, and the exact PT temperature $T^*$ will be determined by the Polaykov loop potential.  Without losing generality, The temperature $T^*$ can range from 100 MeV to a few TeV. Note that if the scalar field is coupled to the SM QCD,  we also need to assume $<\varphi>$ rolls to zero at a temperature below 1 GeV from current QCD constraints to recover the SM QCD if we are modifying the SM sector.
This can be achieved in several ways. For example, we can consider that the non-zero vev of $\phi$ is due to the thermal effect at higher temperatures. The simplest example is constructing the rolling potential e.g.$V(\phi) \sim \phi^2 $ with $\phi$ rolling to zero above 1GeV.
In principle, this scalar can be a second heavy Higgs or inflaton, which will induce a primordial cosmic PT during inflation, with interesting consequences on CMB, which we leave for future work. If we are achieving this PT in the beyond the Standard Model dark sector, known as dark QCD, then the required gauge coupling for the PT to happen can be just put in.

\paragraph*{First order PT of Polyakov loop for $N_c\geq$ 3. }

The chiral PT is closely related to the coupling strength of QCD, while the PT characterized by the Polyakov loop is more sophisticated. It is possibly related to the deconfinement PT. However, here we may refrain from the complexity of this topic, and simply treat it as a PT in the YM sector which is then a transition of the non-Abelian group structure of the states.

The one-loop perturbative computations show that the gluon PT can take place once there exists an effective mass for gluon. Therefore, such a PT is not directly related to the coupling strength of the theory. Note that it still requires a strong coupling to bring in the dynamical mass generation in the conventional  QCD studies with SM. Now as we illustrated here, a new scalar field that coupled with the gluon field in the early cosmological period will also create an effective mass for gluon. It will then induce a new gluon PT alone in the early cosmological period without changing any other phenomena. Assuming there exists a new field that only couples to the non-Abelian gauge field, this PT appears inevitably when the scalar VeV energy scale reaches the new physics scale.

We first take   SU(3) as an example. The Polyakov loop means a nonvanishing value of background gauge field which we denote as $\varphi_3=\beta g \bar{A}_0^3$, and $\varphi_8=\beta g \bar{A}_0^8$.  The one-loop computation of the effective potential with a massive gluon can be expressed as~\cite{Reinosa:2014ooa}:
\begin{align}
&\mathcal{V}=2 T^d(\mathcal{W}_{m_g/T}(0)+\mathcal{W}_{m_g/T}(\varphi_3) \nonumber \\
&+\mathcal{W}_{m_g/T}(\varphi_+)+\mathcal{W}_{m_g/T}(\varphi_-)),
\end{align}
with $\varphi_\pm=(-\varphi_3\pm\sqrt{3}\varphi_8)/2.$ and
\begin{eqnarray}
&&\mathcal{W}_{m_g/T}(\varphi)=\frac{1}{2}[(d-1)F_{m_g/T}(\varphi)-F_0(\varphi)],\\
&&F_{m_g/T}(\varphi)=\int d^{d-1}q \mathrm{ln}(1+e^{-2\epsilon_q}-2e^{-\epsilon_q}cos\varphi),
\end{eqnarray}
with $\epsilon_q=\sqrt{q^2+m_g^2/T^2}$.
There are two minima in the potential $\mathcal{V}$ which implies that it is a first-order PT. Note that here the QCD coupling is not largely enhanced and hence, the quark sector can be still treated perturbatively. Therefore,  there is no dynamical quark mass generation and no related chiral PT to take place.  The center symmetry for the quark sector simply follows the gluon, as the gluon dynamics is dominant at high temperature~\cite{Gale:2012rq,Shuryak:2014zxa}. The QCD phase structure here can be then mainly featured as the PT in the gluon sector characterized by the Polyakov loop. As we will illustrate in the following, for $N_c\geq$ 3, such a PT for $SU(N_c)$  is always first-order.

\begin{figure}[t]
\includegraphics[scale=0.55]{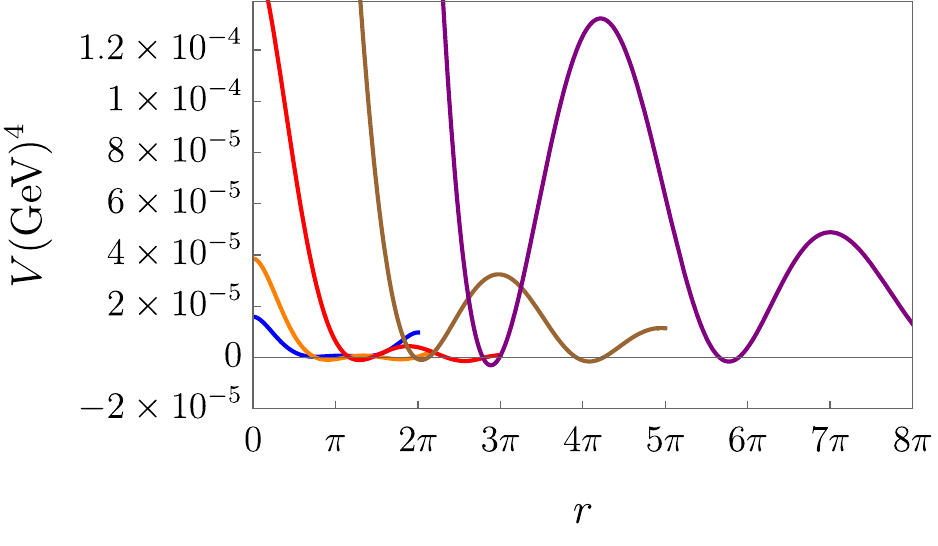}
\caption{
{The effective potential of SU(N) at $T=T_c$ with respect to the main direction $\varphi$, where only the first two minimums are plotted, with $N_c=3,5,10,25,50$ from left to right. }
\label{effectivep} }

\end{figure}

Now one may generalize the above analytical form in the adjoint representation of $SU(N_c)$ theory by analyzing the structure constant. First of all, for $SU(N_c)$, there are $N_c-1$ directions of $\varphi_l$ with  $l\in(2, N_c)$, and the potential can be read as:
\begin{equation}
\mathcal{V}(\varphi_l)=T^d\{(N_c-1)W(0)+2\sum_{m,n} W(\!\!\!\!\sum_{l=m/k/n}\!\!\!\!c_{l} \varphi_{l})\},
\end{equation}
with $1\leq m<k<n\leq N_c$. And for the coefficients $c_l$, one has:
\begin{equation}
c_m=-\sqrt{\frac{m-1}{2m}},\,\,c_n=\sqrt{\frac{n}{2(n-1)}},\,\,c_k=\sqrt{\frac{1}{2k(k-1)}}.
\end{equation}

It is difficult to deal with the $N_c-1$ directions of order parameters, and hence, here we simply define one main direction of r. The strategy is that for each $m$ and $n$,  the variable of the potential can be considered as a $N_c-1$ dimensional vector. and there are $\frac{N_c(N_c-1)}{2}$  vectors in total.  The main direction with one single order parameter $r$ can be then taken as the average of all the vectors which then reads as:
\begin{equation}
\varphi_{l}=\frac{ \sum_{m,n}c_{l=m/k/n}}{\sqrt{\sum_l (\sum_{m,n}c_{l=m/k/n})^2} } \varphi.
\end{equation}
Note that the potential defined here is the same as the location of the minima at $\varphi=\bar{\varphi}$ where $\partial \mathcal{V}/\partial \varphi=0$.  The PT takes place at $T=T_c$ when the pressure of the two phases becomes the same as $\mathcal{V}(\bar{\varphi}_1)=\mathcal{V}(\bar{\varphi}_2)$. The behavior of the potential is depicted in Fig.~\ref{effectivep}. 

Note that for  $N_c=2\,\mathrm{and}\,3$, our potential returns to the expression as shown in Ref.~\cite{Reinosa:2014ooa}, and hence has the same behavior. For $N_c=2$, the transition is second order at $m_g/T_c=2.97$ for $d=4$, and for $N_c=3$ is first order at $m_g/T_c=2.75$. For $N_c\geq 3$, the PT is always first order and $T_c$ remains almost the same as in the case of $N_c=3$.  For large $N_c$, the number of terms in the effective potential increases as $N_c^2$ which then makes the quantities like pressure, energy density, and latent heat to be $N_c^2$ dependence.  These results are all consistent with the studies of lattice QCD simulations~\cite{Lucini:2005vg,Lucini:2012gg,Datta:2010sq}. Moreover, besides of the first two minima, there also exist subleading minima in the effective potential which might indicate the possibility of multi-phase transitions as mentioned in Ref.~\cite{Lucini:2012gg}.  Nevertheless, here we will focus on the first two minima and their related FOPT which can be applied to determine the parameters $\alpha$, $\beta$,$T_*$ in the GW spectrum which will be discussed in the following.

It needs to be mentioned that the PT of the gluon sector at $m_g/T\sim2.75 $ has been found in QCD at around $T=200$ MeV which is caused by the dynamical mass generation of gluon field~~\cite{Lucini:2012gg,Fister:2013bh,Reinosa:2014ooa}. Here either through a similar mechanism in the dark YM sector, or via the scalar field coupled to the QCD gauge sector,  an effective mass of gluon can be generated that brings in a new first-order PT.  One may estimate where a first-order PT appears for the new YM coupling as   \begin{equation}
\alpha_{\mathrm{new}}(T_c) N_c\sim5.45,
\label{PT1}
\end{equation}
 or with  the value of the scalar condensate  coupled to the YM sector :
  \begin{equation}
4\pi\left.\frac{<\phi>}{\Lambda_{pt}}\right |_{T_c}\sim \frac{N_c}{5.45}-\frac{33-2n_f}{12\pi}\mathrm{ln}(\frac{\mu^2}{\Lambda_{QCD}^2}).
\label{PT2}
  \end{equation}
 


 \paragraph{Gravitational waves production}

In contrast to the vev-induced PT, the gluon PT is characterized by the background condensate in the Polyakov loop, which is dynamically generated purely from the nonperturbative dynamics of the Yang-Mills field itself. The details of the spectrum shape can be studied with the effective potential gluons PTs.

There are three contributions to the stochastic gravitational wave background arising from a strong first-order cosmic PT \cite{maggiore,Aggarwal:2020olq,Weir:2017wfa,1990The}:
\begin{equation}\label{eq:total2}
h^2\Omega_{\rm GW} \simeq h^2\Omega_{ \varphi} + h^2\Omega_{\rm sw} +
h^2\Omega_{\rm turb}\,.
\end{equation}
with the contributions of bubble walls, sound waves, and turbulence.

It is widely expected that the sound wave source will dominate and the form of the spectra can be estimated by a broken power law \cite{Caprini:2015zlo,Hindmarsh:2017gnf}
\bea
&h^2\Omega_{\rm \varphi, sw}(f)\nonumber \\
 \sim  10^{-5}
& \left( \frac{H_*}{\beta} \right)^2 
 \left( \frac{100}{g_*} \right)^{\frac{1}{3}} \Upsilon \Gamma ^2 \bar{U}_f^4 v_w S(f) \, ,
\label{eq:Omenv}
\eea
where $\beta/H_\ast$ is the inverse time scale of the transition, normalized to Hubble,
\begin{equation}
\frac{\beta}{H_\ast} = \left. T \frac{d(S_E/T)}{dT} \right|_{T=T_\ast},
\end{equation}
 by solving the classic equation of motion of the order parameter field and calculating the  action at $\varphi=\bar\varphi$ as:
  \begin{equation}
            {{S}_{E}}=4\pi \int_{0}^{\infty }{dr{{r}^{2}}\left[ \frac{1}{2}{{\left( \frac{d\bar\varphi }{dr} \right)}^{2}}+{\mathcal{V}}\left( \bar\varphi  \right) \right]}.
        \end{equation}
The percolation temperature $T_\ast$ is set when the Euclidean action is $S_E\sim 140 T_\ast$. Besides, $\Gamma=4/3$ is the adiabatic index, $\bar{U}_f^2= \frac{3}{4}\kappa_f \alpha $ is the root-mean-square (RMS) fluid velocity, $v_w$ is the wall velocity, $\Upsilon =1-1/\sqrt{1+2(8 \pi ^{1/3}v_w H_{\ast})/(\beta \bar{U}_f)} $ is a suppression factor for the finite lifetime of the source \cite{Guo:2020grp} and $S(f)$ parametrizes the spectral shape,
\begin{equation}
    S_{\rm sw}(f) = \left( \frac{f}{f_{\rm sw}} \right)^3 \left( \frac{7}{4+3(f/f_{\rm sw})^2} \right)^{7/2} \ .
\end{equation}
The peak frequency is controlled by the mean bubble separation which can be related to the thermal parameters we have already introduced
\begin{equation}
    f_{\rm sw} = 8.9 \mu {\rm Hz} \left( \frac{\beta}{h_\ast }\right) \left(\frac{T_{\ast}}{100 {\rm GeV}} \right) \left( \frac{g_\ast}{100} \right)^{1/6} \ .
\end{equation}
Finally, we have defined the fluid velocity in terms of the kinetic energy fraction $\kappa _f$ and the trace anomaly normalized to the radiation energy density \begin{equation}
  \alpha = \left. \frac{\Delta V - \frac{1}{4} T\frac{d\Delta V}{dT}}{\rho _{\rm rad}} \right|_{T=T_\ast} \ .  
\end{equation}

\begin{figure}[t]
\includegraphics[scale=0.5]{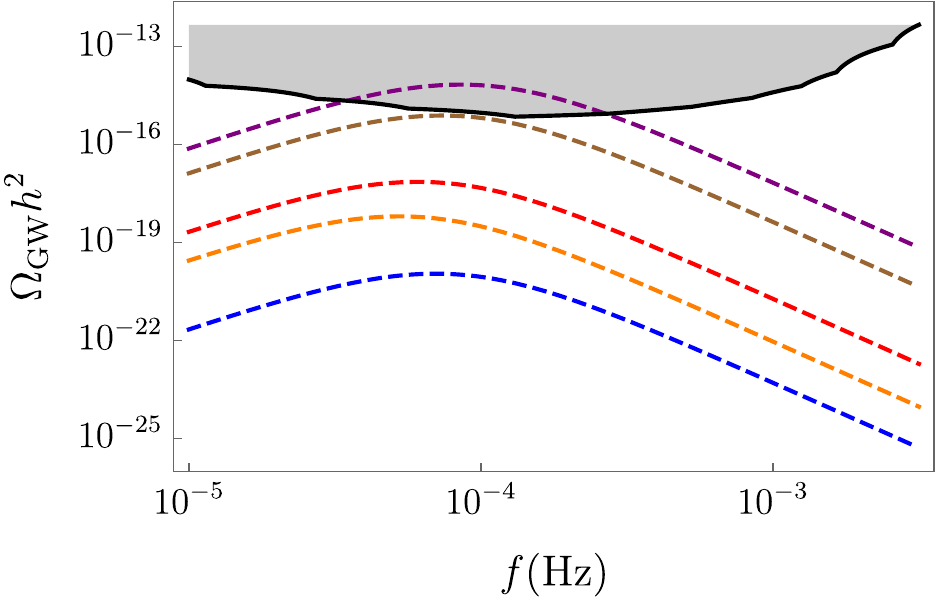}
\caption{
 The gravitational wave power spectrum for a thermal PT 
 at $T_{*}=100$MeV  for the cosideration of $\mu$Ares with $N_c=3,5,10,25,50$, respectively. The temperature can be varied for other GW frequencies. We also plot the sensitivity curves of the proposed space interferometry programs $\mu$-Ares \cite{Sesana:2019vho} in grey. Notice $T_*$/peak frequency can be shifted higher up to multi TeV and the $\Omega_\text{GW}$ will roughly stay the same. In that case, one can reach the PT with different interferometry programs.
\label{fig:gw1}}
\end{figure}

We calculate the Euclidean action by numerically solving the bounce solution to the equation of motion for each value of the temperature assuming a tunneling action from the high-temperature minimum to the nearest neighbor minimum. We also assume a relativistic wall velocity $v_w=1$ and therefore the kinetic energy fraction is given by the detonation regime \cite{Espinosa:2010hh}
\begin{equation}
    \kappa_f = \frac{\alpha}{0.73+0.083 \alpha + \alpha} \ .
\end{equation}

\begin{table}[htbp]
	\centering
	\begin{tabular}{|c|c|c|c|c|c|}
\hline
$N_c$&3&5&10& 25&50\\
\hline
$g_*\alpha$&0.65&2.14& 5.72& 49.76&197.74\\
\hline
$\beta/H $$\times10^{-5}$&1.36&1.02& 1.18& 1.43&1.66\\
\hline
	\end{tabular}
	\caption{$N_c$ dependence of $\alpha$ and $\beta$  }
	\label{tab:para}
\end{table}
We have $\alpha,\beta/H$ roughly staying constant for the range of temperatures we considered. Besides, the fit of data gives that $\beta$ increases mildly as $N^{0.22}_c$ and $\alpha$ increases with $N^2_c$ as depicted in Tab.~\ref{tab:para}. The parameter $\alpha$  directly comes from the trace anomaly and also the latent heat with the same $N_c^2$ dependence, while  $\beta$  is related to surface tension $\sigma$. The power of surface tension can be approximately estimated as    $\sigma\propto(\delta P^2\beta)^{\frac{1}{3}}\sim N_c^{1.4}$. This $N_c$ dependence of surface tension is consistent with the estimate from lattice QCD which varies from $N_c$ to $N_c^2$~\cite{Lucini:2005vg,Lucini:2003zr}. It needs to be mentioned that in Ref.~\cite{Garcia-Bellido:2021zgu}, it has been predicted that   $\beta/H$ shrinks with $N_c$ as  the surface tension  grows with $N^2_c$  from using a combination of lattice data and classical nucleation theory. The method   gives a smaller $\beta/H$  and therefore a stronger PT. However, the results there are with the caveat of applying   the potential  at critical temperature instead of  the nucleation temperature which is quite different based on the effective potential we obtained here.  
 Our final results are presented in Fig.\ref{fig:gw1}. As a benchmark, the peak frequency in \ref{fig:gw1} is set to be $10^{-4}$Hz, although the peak frequency can be shifted, and the strength of gravitational waves $\Omega_\text{GW}$ stays unchanged. 
We can also detect the gluon PT happened above 100 GeV with the future space interferometry programs, e.g. LISA\cite{Barausse:2020rsu}, eLISA\cite{Caprini:2019egz}, Decigo(BBO)\cite{2011CQGra..28i4011K,2006CQGra..23.4887H,Kawamura:2020pcg}, ALIA\cite{Gong:2014mca}, Chinese programs Taiji\cite{Hu:2017mde} and Tianqin\cite{TianQin:2020hid,TianQin:2015yph}. Notice here cosmic PT temperature is determined by Eq.\ref{PT1} or Eq.\ref{PT2}.

\paragraph{Discussion and summary}

The key point of our work is that we illustrate a scenario for the possibility of cosmic gluon first-order PT at the new physics scale.  Due to the extremely high temperature, the quark part of this scenario stays relatively separated in the early universe, and not necessarily strongly coupled.  The phenomenon is purely induced by the gluon PT characterized by the Polyakov loop. Moreover, such a gluon PT is of first order for SU($N_c$) with $N_c\geq 3$ Yang-mills, but second order for SU(2) gauge theory. if one tries a similar coupling to the electroweak gauge bosons, the consequences will be very different. Another plausible way to extend the Standard Model and realize the first-order QCD PT in the early universe is surely by the chiral PT in the quark sector.  However, in a very high temperature with the six flavor quarks being considered to be massless, the PT is very unlikely to be the first-order based on the very recent lattice study. The first-order chiral PT requires additionally the large chemical potential,   which can be achieved by a large lepton asymmetry from beyond the Standard Model extra leptons and/or leptogenesis mechanism. These two PTs, if they exist, happened at relatively separated sectors and different temperatures. They are not mutually exclusive and can be realized together in the early universe. 
Both of these early universe first-order PTs can have imprints on BBN and cosmic evolution.
Our assumption of the rolling scalar and consequently a massive gluon at a very high temperature may overlap with the other new physics phenomena and one can see e.g.\cite{Ipek:2018lhm,Lu:2022yuc,vonHarling:2017yew,Davoudiasl:2019ugw} for more discussions.  

\begin{acknowledgments}
{\bf Acknowledgments:}
FG is supported by the National  Science Foundation of China under Grants  No. 12305134. SCS thanks for the support from the National Natural Science Foundation of China (Nos. 12105013). GW acknowledges the STFC Consolidated
Grant ST/L000296/1 
\end{acknowledgments}

\bibliography{refs.bib}

\end{document}